\documentclass[usenatbib]{nature}

\usepackage[labelfont=bf,textfont=footnotesize,singlelinecheck=on]{caption}

\DeclareCaptionLabelFormat{dash}{#1 #2 -- }
\usepackage{amssymb}
\usepackage{url}
\usepackage{graphicx}
\usepackage{longtable}

\newcommand\teff{T$_{\rm eff}$}
\newcommand{\msun}{\ensuremath{\, {M}_\odot}}
\newcommand{\Msun}{\ensuremath{\, {M}_\odot}}

\newcommand{\wcrit}{$\omega_{\rm crit}$}
\newcommand{\win}{$\omega_{\rm in}$}

\newcommand{\xc}{X$_{\rm c}$}
\newcommand{\Tc}{T$_{\rm c}$}

\newcommand{\mcore}{M$_{\rm core}$}
\newcommand{\mir}{m$_{\rm F814W}$}

\newcommand{\finf}{m$_{\rm F814W}$}
\newcommand{\coluv}{m$_{\rm F336W}$--m$_{\rm F814W}$}
 
\newcommand\aj{{Astron.~J}}%   
\newcommand\apj{{Astrophys.~J}}%   
\newcommand\apjl{{Astrophys.~J.~Letters}}%   
\newcommand\apss{{Astrophys.~Space~Sci.}}%   
\newcommand\aap{{Astron.~Astrophys.}}%   
\newcommand\aapr{{Astron.~Astrophys.~Rev.}}%   
\newcommand\mnras{{Mon.~Not.~R.~Astron.~Soc.}}%   
\newcommand\nat{{Nature}}%   
\usepackage[usenames,dvipsnames]{color}

\title{Stars caught in the braking stage in young Magellanic Clouds clusters
}
\author{Francesca D'Antona$^{1}$, Antonino P. Milone$^{2}$,  Marco Tailo$^{3}$, Paolo Ventura$^{1}$,   Enrico Vesperini$^{4}$,  Marcella Di Criscienzo$^{1}$ }
\begin{document}

\maketitle

%\begin{affiliations}
 %\item Put institutions in this environment and
 %\item separate with \verb|\item| commands.
%\end{affiliations}

%================================================================
\begin{affiliations}
 \item INAF- Osservatorio Astronomico di Roma, I-00040 Monte Porzio (Roma), Italy.
 \item Research School of Astronomy \& Astrophysics, Australian National University, Canberra ACT 2611, Australia
 \item Dipartimento di Fisica, Universit`a degli Studi di Cagliari, SP Monserrato-Sestu km 0.7, 09042 Monserrato, Italy
 \item Department of Astronomy, Indiana University, Bloomington, IN (USA)
 %\item Space Telescope Science Institute, 3700 San Martin Dr., Baltimore, MD 21218, USA
 % \item INAF- Osservatorio Astronomico di Bologna, via Ranzani 1, I-40127 Bologna, Italy
 %\item INAF, IAPS, Roma, via Fosso del Cavaliere 100, I-00133 Roma, Italy
\end{affiliations}
%================================================================

%%%%%%%%%%%%%%%%%
%%%%%%% sono 210 parole. Togliere 10 parole!!!%%%%
%%%%%%%%%%%%%%%%%
\begin{abstract}
%For Nature, the abstract is really an introductory paragraph set in bold type.  This paragraph must be ``fully referenced'' and less than 180 words for Letters.  VERAMENTE DICE 200- This is the thing that is supposed to be aimed at people from other disciplines and is arguably the most important part to getting your paper past the editors.  End this paragraph with a sentence like ``Here we show...'' or something similar.
%SONO 237 PAROLE PER ORA
% 222 dopo aver eliminato un po' di bibliografie "flourishing"
%%%%%%%%%% 205 finale (meno "NEW" sono 204) 
 The color-magnitude diagrams of many Magellanic Cloud clusters (with ages up to 2 billion years) display extended turnoff regions where the stars leave the main sequence, suggesting the presence of multiple stellar populations with ages which may differ even by hundreds million years\cite{mackey2008,milone2009, girardi2011}. A strongly debated question is whether such an extended turnoff is instead due to populations with different stellar rotations\cite{girardi2011, goudfrooij2011, rubele2013, li2014natur}.
The recent discovery of a `split' main sequence in some younger clusters ($\sim$ 80--400\,Myr) added another piece to this puzzle. The blue (red) side of the main sequence is consistent with slowly (rapidly) rotating stellar models\cite{dantona2015,milone20161755,correnti2017,milone2016sub}, but a complete theoretical characterization of the observed color-magnitude diagram appeared to require also an age spread\cite{correnti2017}. We show here that, in three clusters so far analyzed, if the blue main sequence stars are  interpreted with models that have been always slowly rotating, they must be $\sim$30\%  younger than the rest of the cluster. If they are instead interpreted as stars initially rapidly rotating, but that have later slowed down, the age difference disappears, and ``braking" also helps to explain the apparent age differences of the extended turnoff. The age spreads in Magellanic Cloud clusters are a manifestation of rotational stellar evolution. Observational tests are suggested.
\end{abstract}

%Then the body of the main text appears after the intro paragraph. 
%Figure environments can be left in place in the document. \verb|\includegraphics| commands are ignored since Nature wants the figures sent as separate files and the captions are automatically moved to the end of the document (they are printed out with the \verb|\end{document}| command. However, tables must e manually moved to the end of the document, after the addendum.

When HST observations of the Magellanic Cloud cluster NGC 1856 extended the color baseline from UV to near IR, they revealed the presence of a split main sequence. This feature could not be ascribed to age or metallicity differences\cite{milone2015}, and was not even compatible with a \textit{spread} of rotation rates, but it could be well understood by assuming the presence of two coeval populations: $\sim$65\% of {\it rapidly rotating} ``redder" stars, and $\sim$35\% of ``bluer" non--rotating or slowly rotating stars, evolving off the blue main sequence at a turnoff luminosity  lower than that of the rotating population\cite{dantona2015}, as expected from the results of Geneva tracks and isochrones computations\cite{georgy2014popsint}.
In coeval populations, the evolution is faster (and the turnoff less luminous), for stars with slower rotation rate. In rotating models, the changes due to nuclear burning and rotational evolution are intertwined, as the transport of angular momentum through the stellar layers is associated with chemical mixing by which the convective H--burning core gathers H-rich matter from the surrounding layers, extending the main sequence lifetime\cite{meynetmaeder2000, ekstrom2012, georgy2013a}.
Stars with the same mass but different rotation rates have different evolutionary times and different turnoff luminosities;
this effect helps to produce an extended main sequence turnoff (eMSTO)\cite{niederhofer2015}, but may be insufficient to explain the whole spread observed\cite{correnti2017}.

 Figure 1 shows the HST color-magnitude diagram in the plane of the near infrared magnitude \finf\ versus the  color \coluv, for NGC\,1856 ($\sim$400\,Myr, panel d) and for three younger LMC clusters: (a): NGC\,1755 ($\sim$80\,Myr)\cite{milone20161755}; (b) NGC\,1850\cite{correnti2017} ($\sim$100\,Myr); (c) NGC\,1866\cite{milone2016sub} ($\sim$ 200\,Myr).  In all cases, a split main sequence is present. The interpretation of this split in terms of stellar rotation requires the rotation distribution to be bimodal\cite{dantona2015, milone20161755,correnti2017}, and much more skewed towards high rotation rates than in the field of the Galaxy and of the LMC\cite{dufton2013}  or in low-mass Galactic open clusters\cite{huang2010} possibly suggesting an environmental effect\cite{bastian2017}.

The split finishes at magnitudes corresponding to a kink in the main sequence, due to the onset of surface turbulence at \teff$\lesssim$7000\,K\cite{dantona2002}, where rotational evolution begins to be dominated by the presence of the convective surface layers.
Remarkably,  Figure 1 shows for the first time that {\it in all the three younger clusters a coeval  slow--rotating population does not adequately fit the color-magnitude diagram: the blue main sequence is populated beyond the coeval non-rotating turnoff} by stars resembling the ``blue stragglers" present in some standard massive clusters (e.g. in the old galactic globular clusters\cite{ferraro2009}). These stars can only be explained with {\it younger } non-rotating isochrones, at least $\sim$25\% younger, according to the orange  dashed isochrones plotted in Figure 1. 

We made detailed simulations (see Methods) of the color-magnitude diagrams, excluding NGC\,1856, because the possibly-younger blue main sequence stars are not well distinct from the turnoff stars.
Simulations can not reproduce the brighter part of the blue main sequence with a coeval ensemble of rotating and non rotating stars.  In NGC 1755 (Supplementary Figure 1), inclusion of stars on a younger isochrone can reproduce better the entire non rotating sample and account for the blue main sequence at $18 \lesssim $m$_{F814W}  \lesssim $19. In the other clusters (Supplementary Figures 2 and 3)  the simulations require both a {\it younger} blue main sequence and the presence of stars on {\it older} isochrones (see Supplementary Table 1 and Supplementary Figure 4) to match stars in the extended turnoffs, although the effect of non--sphericity (limb and gravity darkening) and random orientiation accounts for a part of the spread  in the case of high rotation (Supplementary Figure 4).

Although, as discussed later, the fraction of   `younger' stars is  only 10 --15\%, understanding the origin of this population is crucial. A younger rapidly--rotating component would be easily revealed (Supplementary Figure 5), so why does the `younger' population include only slow or non--rotating stars? 

{\it We show here that these stars may represent a fraction of the initially rapidly-rotating stars that have been recently braked: they are not younger in age, but simply in a younger (less advanced) nuclear burning stage.}

The evolution of the core mass (\mcore) and central temperature (\Tc), as a function of the core hydrogen content (\xc), is very similar for non--rotating and rotating tracks (Supplementary Figure 6).  The main difference is the total time spent along the evolution, because, in rotating stars,  mixing feeds the convective core with fresh hydrogen-rich matter and thus extends the main sequence life at each given \xc.  
Therefore, a transition from fast to slow rotation does not require a dramatic readjustment of the star interior.  \\
The angular momentum of the star may be subject to external, additional sinks, besides those included in the models.  It is possible that the external layers are the first to brake (for example if they are subject to magnetic wind braking, as observed in the magnetic star $\sigma$Ori\,B, whose rotation period increases on a timescale of 1.34\,Myr, ref. \cite{townsend2010}) and that the information propagates into the star by efficient angular momentum transport. 
Otherwise, braking first occurs into the core, e.g. by action of  low-frequency oscillation modes, excited by the periodic tidal potential in binary stars  (dynamical tide\cite{kopal1968,zahn1977}), as proposed\cite{dantona2015} for the case of NGC\,1856. We prefer the latter mechanism, as its timescale depends both on the stellar mass and its evolutionary stage, and on the parameters of the binary system, so braked stars (the upper blue main sequence) may be present in clusters over a wide range of ages. The stellar envelope will be the last to brake, and then the star will finally reach the location of the non-rotating configuration on the colour magnitude diagram. If this latter stage takes place before the end of the main sequence phase, the star will be placed on the blue main sequence.
A star moving from the rotating to the non--rotating evolutionary track {\it at fixed \xc} will appear {\it younger}
as soon as braked, while its total main sequence time will be {\it shorter} --simulating an older isochrone-- than that of the star preserving its rotation rate, because full braking may prevent further core--envelope mixing. This produces two different effects: the presence of a {\it younger} blue main sequence, and of {\it older} stars showing up in the puzzling extended turnoff.

Both these `age' effects are schematically illustrated in \textbf{Figure 2}. 
%%%%% FRANCA: servirebbe, ma non c'è spazio? RIMESSO, scrivere all'Editor
Here we must keep in mind that the explanation is drawn based on existing stellar models, and a strong computational effort will be needed in future to confirm this suggestion.
For the clusters NGC 1755 (left panels) and NGC 1866 (right), we plot the time evolution of core hydrogen \xc(t) for selected
masses, for models initially rotating with \win=0.9\wcrit\ (where \wcrit\ is the breakup angular velocity required for the centrifugal force to counterbalance gravity at the equator) and for non-rotating models (\win= 0).
A vertical line marks the location of these masses at the age of each cluster, where we see that the rapidly rotating star is in a less advanced nuclear burning stage.
The grey arrows shows the age of a star born non rotating having the same \xc. ``Rapid braking" would shift the star to the non-rotating radius (and color-magnitude location) corresponding to that same \xc, so it would appear to us ``younger" than a star with the same mass but formed with no rotation (blue squares). In fact (panels b and d) the braked masses will be approximately located on the mass--\xc, \win= 0 isochrone (open green dots on the green lines) 25\% younger than the \win= 0 isochrone at the clusters' age (blue dashed, where the stars non-rotating from the beginning --blue squares-- are placed).  The presence of stars on a ``younger" blue main sequence can thus be qualitatively understood. These stars must have been fully braked ``recently", less than 25\% of the cluster age ago, otherwise they would have already evolved out of the main sequence. 

A second consequence of the braking process is that the time evolution \xc(t) of each ``braked" star will depend on the time at which braking is effective in changing the modalities of mixing at the border of the convective core to the \win= 0 modality.
Simplifying, the braked stars stop evolving along the \win= 0.9 \wcrit\ \xc(t), and start evolving along the \win= 0 \xc(t), at different times (dashed grey lines in a and c of Figure 2, see Methods). The intersections of the dashed grey lines with the vertical line drawn at the cluster age show that each mass may, in principle, span the whole range of  \xc\ between the minimum value achieved by the  non-rotating track and the maximum value of the rotating track. 

Braking will be in reality much more complex than this exploratory outline.  Two possibilities can help the interpretation of the color-magnitude diagram patterns: 

1) The mechanism for slowing down the stellar core might cause strong shear in the outer layers, and  imply {\it even more mixing} than in the standard rapidly rotating models, before the star is finally fully braked. A more extended mixing explains why part of the upper blue main sequence stars in NGC\,1866 and NGC\,1850 look younger than predicted by the $\sim 25$\% difference between the rapidly rotating and the non rotating isochrones (see Supplementary Table 1 and Supplementary Figure 4).

2) Full braking of the external layers (corresponding to the blue main sequence stage) is possibly achieved by only a fraction of braking stars, and the ``older" stars of the extended turnoff may be directly evolving  from the rotating main sequence and not from the blue main sequence.  In Supplementary Figure 2 and 3 we simulate the dimmer extended turnoff's of NGC\,1866 by samples of stars extracted from rapidly rotating older isochrones (see also Supplementary Figure 4). In fact, the number versus magnitude plot of the blue main sequence stars is practically flat   until  \mir$\lesssim$20--20.5, whereas number counts of the rest of the stars increase, as expected for any standard mass function (Supplementary Figure 7). Thus the fully braked stars seem not to "pile up" on the blue main sequence, not even at magnitudes where we should see stars braked for the whole cluster lifetime. This may indicate that also stars much dimmer than the blue main sequence turnoff have reached full braking only recently, and that the aging effect of braking is seen mainly in the extended dimmer portions of the rotating turnoff. Piling up of slowly-rotating stars braked at different ages produces a significantly populated turnoff of the blue main sequence only at the age of NGC\,1856\cite{dantona2015}, but it is not evident in the younger clusters.
 
These initial results may shed some light on the physical mechanism behind the braking. As both the ``blue stragglers" and the extended turnoff require braking in recent times, does braking accelerate for stars already in advanced core hydrogen burning?   
In the dynamical tide mechanism, the synchronization time increases with the age of the binary system\cite{zahn1977,zahn2008}, but we can expect that the detailed behavior of angular momentum transfer  and chemical mixing at the edge of the convective core is more subject to small differences in the parameters when the structure is altered by expansion of the envelope and contraction of the core. 
In addition, the timescale will depend on parameters which may vary from cluster to cluster, possibly including the location of the star within the cluster. For instance, the blue main sequence fraction increases in the external parts of NGC\,1866\cite{milone2016sub}, while it does not vary with the distance from the cluster center in other clusters\cite{li18562016,correnti2017}. 

We conclude that rotational evolution produces different timescales for the core--H burning phase which can be perceived as a mixture of stellar ages. The most direct indication in support of this interpretation comes from the presence of a small population of non--rotating stars which appear to be {\it younger} than the bulk of stars. Stars whose envelope is  not fully braked may instead show up as {\it older} and be seen in the dimmer part of the extended turnoff.
``Complete" simulations of the color-magnitude diagram (see Supplementary Table 1) requires an age choice for the bulk of stars (the age defined by the most luminous rapidly rotating population) plus smaller samples of different ages (Supplementary Figure 4), younger, to describe the blue main sequence, and older to describe the multiple turnoff (Supplementary Figures 1, 2 and 3), but all the stars may in fact be a coeval ensemble.

The best test for the model of the blue and red main sequence in terms of different rotation rate will be to find low rotation velocities in the spectra of the blue main sequence stars, and larger velocities in the red main sequence stars. 
The H$\alpha$\ emission typical of rapidly rotating stars (Be stage) should be  mostly confined to the red turnoff stars (as in NGC\,1850\cite{bastian2017}).
A test for the braking model is possible by studying the surface anomalies of CNO elements. If the blue main sequence stars are born non rotating, we should expect CNO difference between the blue and red main sequence spectra\cite{ekstrom2012},  but the signatures of CNO cycling will be {\it similar} both for the red and blue side, if the blue side stars have been braked.
For the range of masses evolving in the studied clusters, rotational mixing increases the helium content only marginally at the surface, but the ratio N14/C12 increases by a factor $\sim$4 at 3.5\Msun (NGC\,1866) and $\sim$5 at 5\msun (NGC 1755 and 1850), with respect to the non-rotating counterparts. In younger clusters showing split main sequence, both helium and the N14/C12 ratio would be more affected. \\
%%%%%%%%% ADDED: signal to the referee N.1#########
Finally, notice that, if the braking is due to a tidal torque, the upper blue main sequence stars should have binary companions. A full exploration of the binary properties of the blue and red turnoff stars may shed further light on the evolution.

{\begin{figure}
%\vskip -100pt  
	\includegraphics[width=16cm]{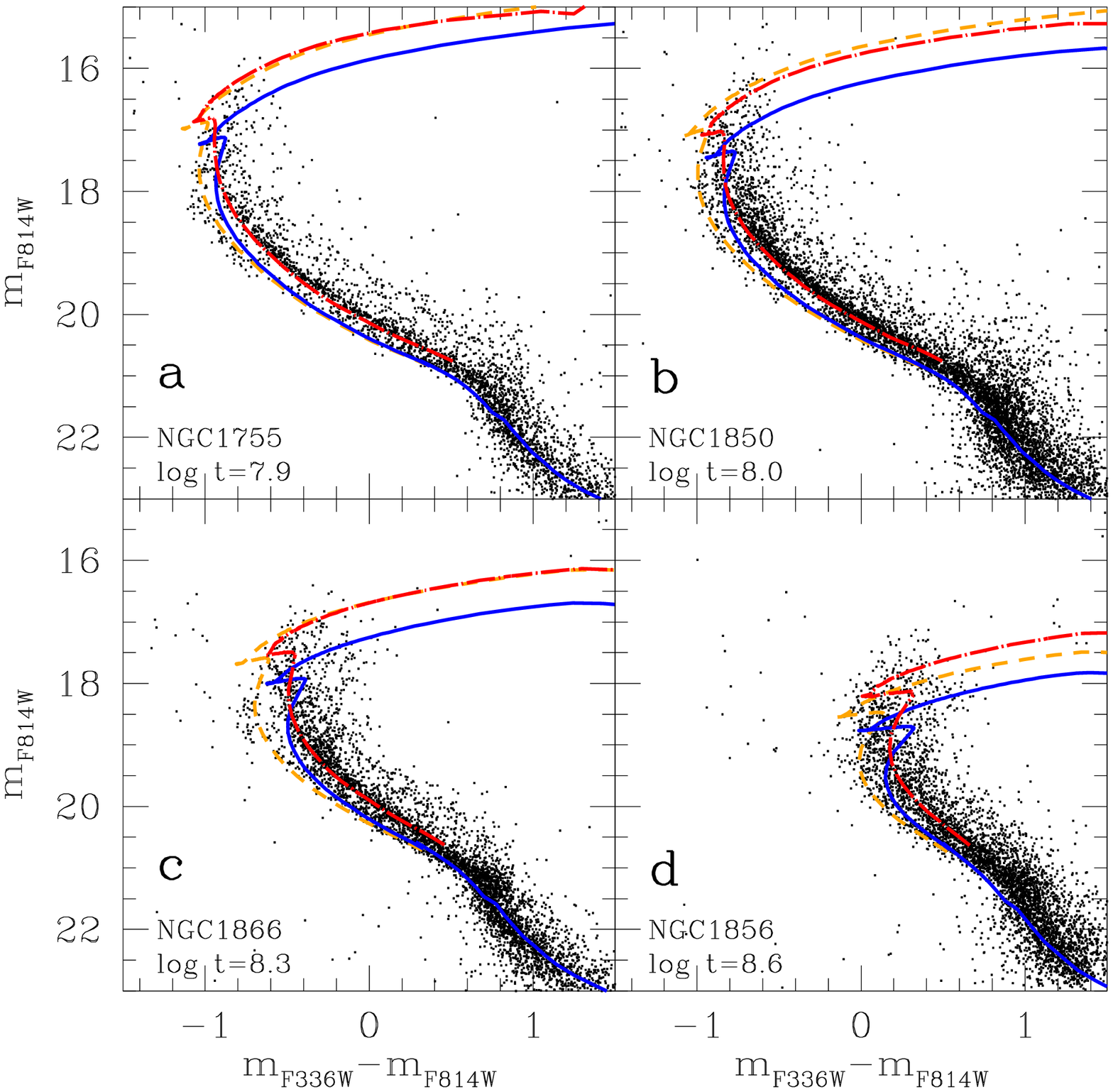}
%\vskip -100pt  
\caption{\textbf{Color-magnitude diagram for four young LMC clusters at different ages.} \\ 
At the bottom of panels from a to d, the clusters are identified, and the adopted logarithm of the age (in years) is labelled. All diagrams are characterized by an evident  split of the MS, although the split extent in magnitude decreases as the cluster age increases. Coeval isochrones for \win=0 (blue, solid) and \win=0.9\wcrit\ (red, long dash--dotted), where \wcrit\ is the break up angular velocity, are shown. Orange dashed lines are \win=0 isochrones younger by 0.1\,dex than ages labelled at bottom.  Younger slow--rotating isochrones are apparently needed to account for the blue upper main sequence stars. See text for details. }
\label{fig:1}       
\vskip 280pt  
\end{figure}

\begin{figure}
%\vskip -90pt  
	\includegraphics[width=8.cm]{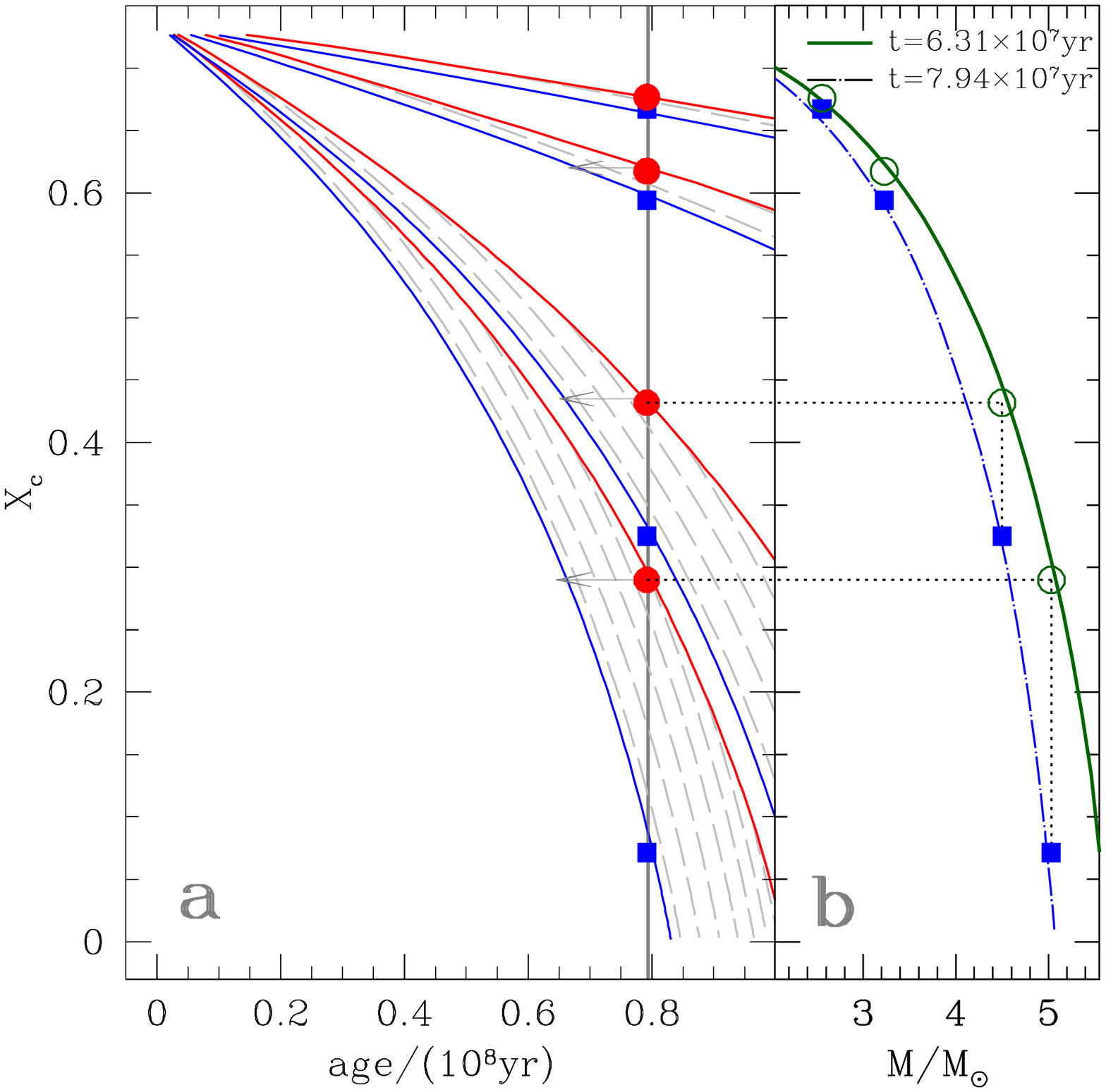}
	\includegraphics[width=8.cm]{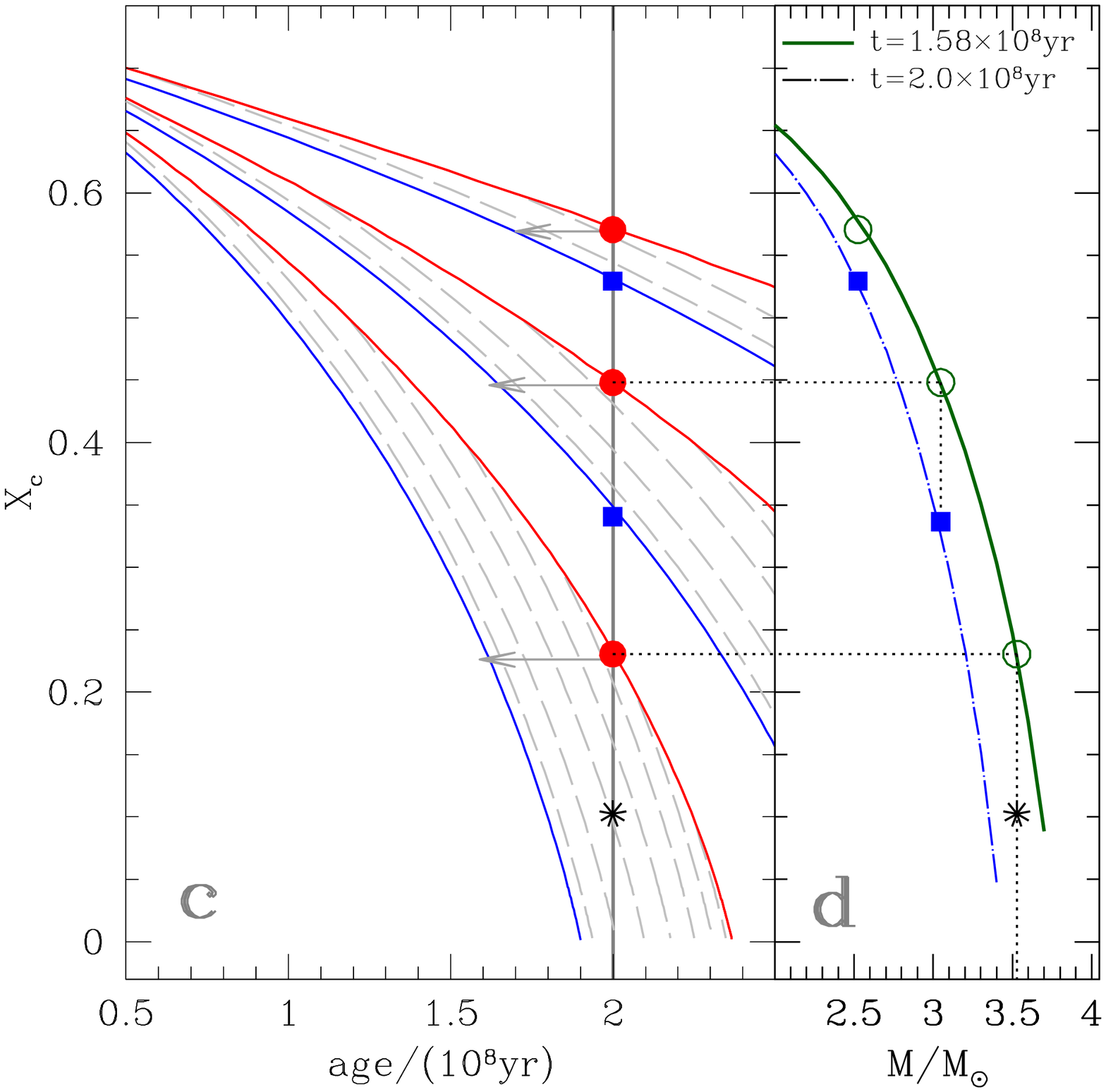}
%\vskip -90pt  
	\includegraphics[width=8.cm]{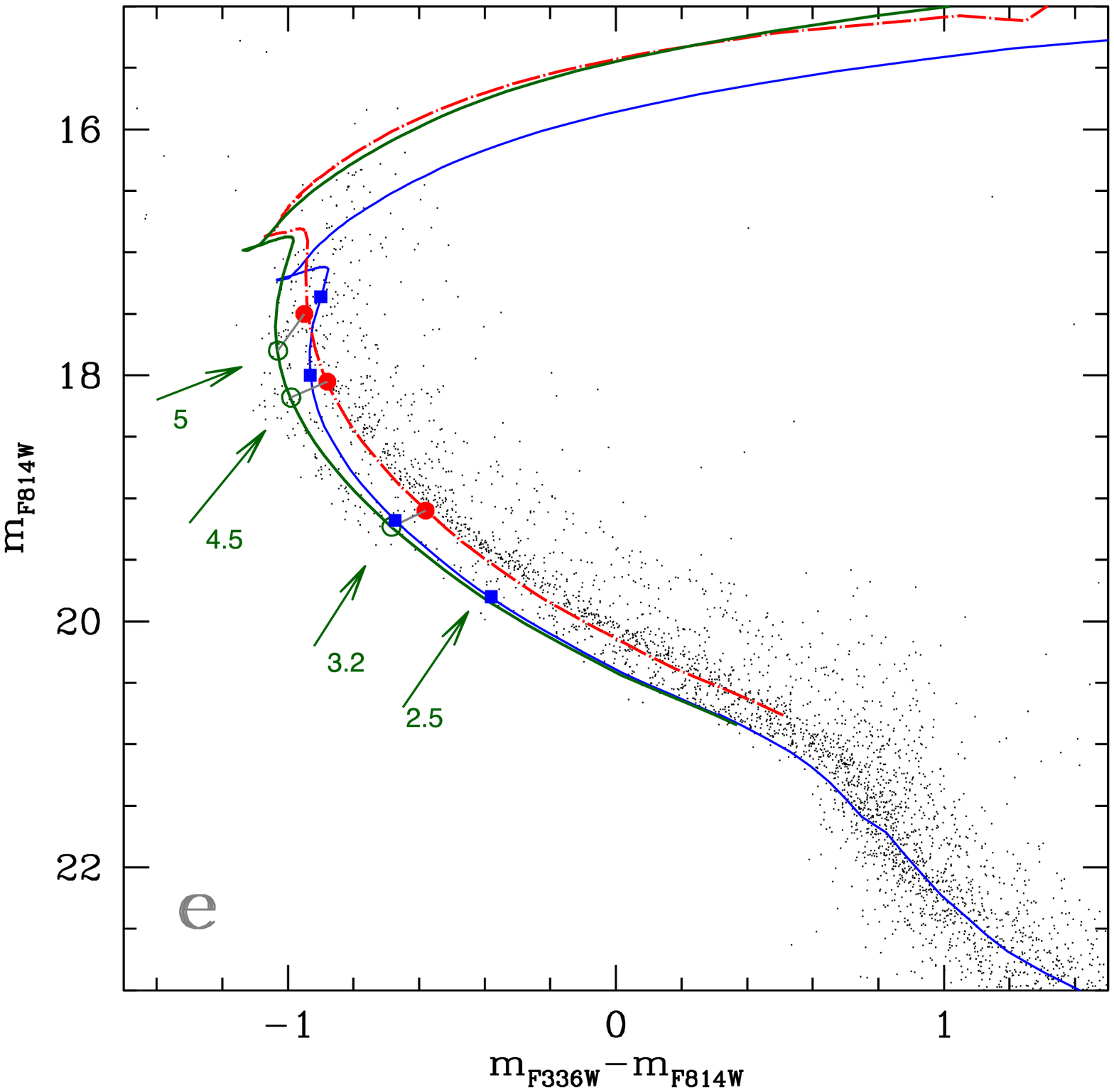}
	\includegraphics[width=8.cm]{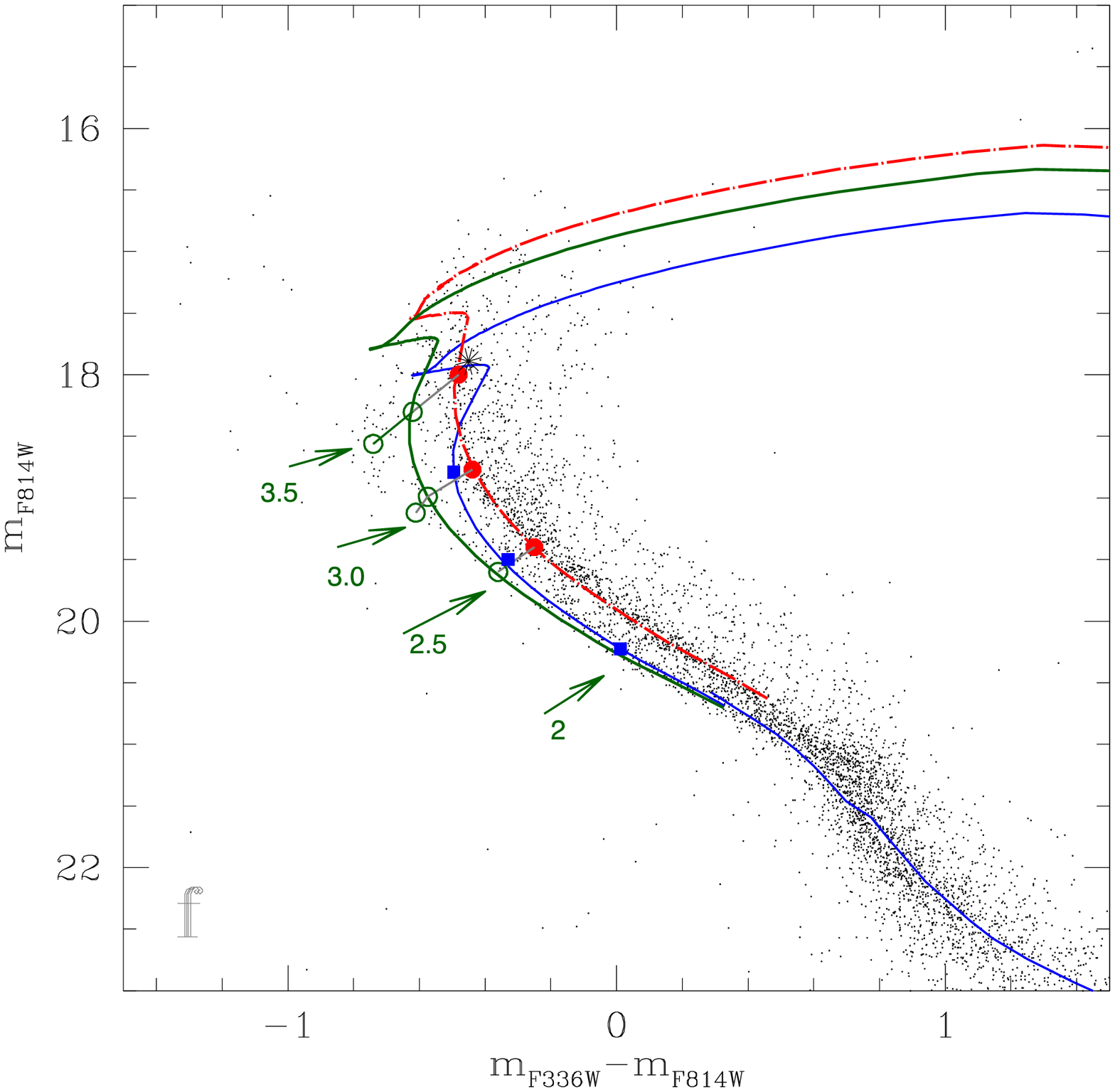}
%\vskip -70pt  
\caption{
\textbf{Illustration of the role of braking at the age of NGC\,1755 and NGC\,1866.}\\
NGC\,1755  is studied in the panels a, b and e, and NGC\,1866 in the panels c, d and f. Panels e and f show the observed data, the isochrones at the cluster age (blue and red), and the isochrone 0.1\,dex younger (green), on which the mass points corresponding to the \xc(t) evolution of the panels a,b and c,d are highlighted.
Panels a and c: Core hydrogen content \xc\ as function of time in units of 100\,Myr, for different masses. For each mass, the upper line (red) is the \win=0.9\wcrit, where \wcrit\ is the break up angular velocity; the lower line (blue) is the \win=0 evolution.  The nuclear burning stage reached at the age of the clusters is marked by red dots (blue squares) for the rotating (non rotating) stage; dark green open dots are the recently braked stars locations, in the plane \xc\ versus mass (panels b and d). The corresponding locations in the color-magnitude diagrams are shown in panels e (NGC 1755) and f (NGC1866), where the mass is labelled in green, in solar units. The dashed grey lines represent schematic transition from \win=0.9\wcrit\ to \win=0, occurring at different ages. For further details, see text. As an example, the asterisks in panels c and d mark the evolutionary stage of a star that braked about 70 Myr ago, so that it is now evolving past the turnoff (asterisk in panel f).}
\label{figure2}       
\vskip 170pt  
\end{figure}
}

\vskip 100pt
\noindent {\bf METHODS}

\subsection{The data sets.}
To study multiple populations in NGC\,1856, NGC\,1755  and NGC\,1866 we have used the photometric catalogs published in our previous papers\cite{milone2015, milone20161755, milone2016sub} and obtained from images collected through the F336W and F814W bands of the Ultraviolet and Visual Channel of the Wide-Field Camera 3 (UVIS/WFC3) on board of {\it HST}. The quoted references provide details on the data and the data reduction.\\
For NGC\,1850 we have used five UVIS/WFC3 images in F336W (durations: 260s, 370s, 600s, 650s, 675s) from GO14069 (PI N. Bastian) and three in F814W (7s, 350s, 440s) from GO14147 (PI. P. Goudfrooij).
 These images have been reduced by adopting the methods developed by Jay Anderson\cite{anderson2008} and used in our works on NGC\,1755, NGC\,1856, and NGC\,1866.
 Photometry has been corrected for differential reddening and small variations of the photometric zero point\cite{milone2012} and has been calibrated to the Vega system\cite{bedin2005} and using the zero points provided by the STScI official webpage.\\
In order to reduce the contamination from field stars we have only used stars in a small region centered on the cluster and with radius of 40 arcsec. In the case of NGC\,1850 we have minimized contamination from the nearby star cluster NGC\,1850B by excluding stars with distance smaller than 20 arcsec from its center. 

\subsection{Models and simulations} 

We make use of the tool SYCLIST (for SYnthetic CLusters, Isochrones, and Stellar Tracks), web facility available at http://obswww.unige.ch/Recherche/evoldb/index/ and created by C. Georgy and S. Ekstr\"om\cite{georgy2014popsint}, both for the stellar models and isochrones.  Details of the physical treatment are contained in the relevant papers of this group\cite{ekstrom2012,georgy2013a}. The models with mass fraction of helium Y=0.26, metals Z=0.006, and $\alpha$-elements in the solar ratios are used, as this composition is the best available to study the LMC clusters.
The models are available for any choice of the initial angular velocity \win\, from 0 to the break--up  \wcrit=$\sqrt{(GM/R_{e,crit}^3)}$, where $R_{e,crit}$ is the equatorial radius at \wcrit.  
The mixing efficiency\cite{ekstrom2012} depends on an effective diffusion coefficient,  accounting both for the meridional circulation and  horizontal turbulence\cite{chaboyerzahn1992} and for the shear--mixing diffusion coefficient. Both radiative and mechanical (equatorial) mass loss are accounted for\cite{ekstrom2012}.
\\
We use for this work plainly the  \win=0.9\wcrit\ models and the \win=0 isochrones, as they account for the color separation of the blue (identified with the \win=0 models) and red (corresponding to the \win=0.9\wcrit\ models) main sequences in the four clusters of Figure 1\cite{dantona2015, milone20161755, milone2016sub}.  A previous analysis of the NGC\,1850 data\cite{correnti2017}  shows that it is necessary to exclude stars with intermediate rotation $0.5 < \omega$/\wcrit$< 0.9$, to let the main sequence remain split. The rotational distribution of field stars\cite{zorec2012, dufton2013}, is much more continuous,  although it also shows signs of bimodality. A hypothesis is that all stars are born rapidly rotating\cite{dantona2015}. It is preliminary confirmed by the presence of the Be type stars, which are all rapidly rotating\cite{rivinius2013}, in the red side in NGC\,1850\cite{bastian2017}. Rotational braking may be due to a non--close companion (orbital period between 4 and 500\,d), as these binaries, in the field, have rotational velocity significantly smaller than for single stars\cite{abt2004}, with about one-third to two- third of their angular momentum being lost, presumably, by tidal interactions\cite{zahn1977}. 
\\
We use the simulations provided by the SYCLIST facility for the rotating sample. This is mandatory for the high \win\ simulations, because they account for the effect of gravity darkening (due to the fact that poles in a rotating star are hotter and brighter than the equatorial region\cite{espinosa2011}) and limb darkening (due to the optical thickness of the atmosphere towards the central and periphery regions\cite{claret2000})\cite{georgy2014popsint}.  Both effects are such that a star seen pole--on will look hotter and more luminous (bluer and brighter) than it would be if seen equator--on. Therefore, the angle of view under which we observe a star will influence its location in the color-magnitude diagram. 
By using a random viewing angle distribution of rotation axes, most of the effect results in a spread in color and luminosity the turnoff region, in agreement with the observations. In Supplementary Figure 4, for the case of NGC\,1866, we compare the simulation from the same \win=0.9\wcrit\ isochrone in which the projection effect is included (yellow triangles) or not (violet triangles) to show this dramatic effect on the turnoff. We note anyway that the color spread is not fully accounted (see ref. \cite{dantona2015} for an extended discussion). In Supplementary Figure 4  we show that the whole turnoff spread is matched by adding  rotating stars at ages 0.05\,dex (red squares) and 0.1\,dex larger (green squares).
\\
The initial mass function in the Geneva database is fixed to Salpeter's\cite{salpeter1955} power-law function with an index $\alpha$=--2.35. We  produced  \win=0 simulations with  smaller values of $\alpha$=--1.0, to fit better the blue main sequence (see the discussion on number counts as a function of m$_{814W}$). 
\\
The data of all synthetic simulation were transformed into the observational planes using model atmospheres\cite{castelli2003}, convolved with the HST filter transmission curves. 
The points were reported to the observational planes by assuming the following color shift  $\delta$c=$\delta$[ m$_{F336W}$--m$_{F814W}$] and distance moduli d=m$_{F814W}$--M$_{F814W}$: 
NGC\,1755   $\delta$c= 0.37\,mag, d=18.50\,mag; 
NGC\,1850:      $\delta$c= 0.52\,mag, d=18.70\,mag;
NGC\,1866:     $\delta$c= 0.32\,mag,    d=18.50\,mag.\\
In the simulations we take into account the combined photometry for a variable percentage of binaries\cite{milone2009}. Binaries are included both in the rotating and non rotating group. The mass function of the secondary stars is randomly extracted from a Salpeter's mass function, with a lower limit of 0.5\msun. Only the photometric consequences of the presence of such binaries are monitored in the simulations.

\subsection{Simulations for the three younger clusters: necessity of a `younger' blue main sequence }

We plot in Supplementary Figures 1, 2 and 3 the Hess diagram of data (left), best simulation (center) and a
simulation using only two coeval (rotating and non rotating) isochrones (right), for the clusters NGC\,1755, NGC\,1850 and NGC\,1866. Supplementary Table 1 lists the relative fraction of samples for the best simulation of the three clusters. 
The right panels are shown to see that the lack of a sample of younger non--rotating stars does not allow to account for the morphology of the color-magnitude diagram. The very evident discrepancy, when a unique coeval non--rotating population is assumed, confirms that it is not possible that rotating
and non-rotating stars born at the same time account for the results. \\
The best simulation is obtained first by choosing the samples of single--age synthetic clusters which best map  the colour magnitude patterns. This is shown in Supplementary Figure 4, where the different colors mark the choices for NGC\,1866. Another example is given in Supplementary Figure 5, where we show in yellow the pattern of the simulation of the \win=0.9\wcrit\ sample which defines the red turnoff stars (and, together with a dating for the cluster, establishes the distance modulus and reddening). We add synthetic populations based on \win=0.9\wcrit\ isochrones {\it younger} than the yellow dots, and see that these points do not correspond to cluster stars locations, so we skip them from consideration. 
Following this first choice, we base the quantitative comparison on number counts.  We first ``rectify" the MS of the three clusters\cite{milone2012}, which helps to separate the blue and red side of the sequences. We then calculate the number of stars in each of the bins described in the first column of Supplementary Table 2 for both observational sequences. We associate to each bin count N a poissonian error {$\sigma = \sqrt{\rm (N)}$. We repeat the procedure for the synthetic clusters and choose the combination of parameters (Supplementary Table 1) that gave us the minimum discrepancy in the ratio between the observed and theoretical sequences.
An outcome of this procedure is to find the population fraction of blue main sequence stars. Examples are shown in Supplementary Figure 7, while Supplementary Table 2 lists the results for each cluster.  The error value associated to the ratio in each bin is calculated from Poisson statistics (where $\sigma = \sqrt(N)$ ) and through standard error propagation procedures. 
 
\subsection{Results and comparisons: how to justify a `braking' track--shift}

In this work, we make a simplified hypothesis: we assume that a rapid braking of the stellar layers may shift the evolution from the \win=0.9\wcrit\ track to the \win=0 track for the same mass. This naive assumption relies on the comparison of the different evolutionary paths shown in Supplementary Figure 6. The \win=0.9\wcrit\ tracks are the lines with squares, and non rotating tracks are simple lines. We show the mass   3.5 and 6\msun. As a function of the core  hydrogen content \xc, we show the age (panel a), the convective core mass \mcore (b), \Tc\ (c) and luminosity (d). The range of values of the physical quantities differ for the different masses, but the behavior is similar. 
Only the \mcore\ evolution is slightly different in the first phases of core--H burning. In fact, its size relies on two counteracting physical processes linked to rotation. In the first phase, rotation generates an additional support against gravity due to the centrifugal force, so the rotating core and the stellar luminosity are smaller.  In the following evolution, the rotational mixing at the edge of the core progressively brings fresh material into the core, increasing its mass, hence its luminosity. The resulting time evolution is very different: in the rotating tracks the total main sequence phase lasts about 25\% longer, despite their luminosity is larger. Thus, at a fixed cluster age, fast rotating evolving stars have {\it larger} masses than the non rotating ones (e.g. at t=100\,Myr, the turnoff masses are $\sim$4.6 and $\sim$4\msun\ respectively). 
\\
If the rotational mixing stops due to external factors which cause braking, the fact that \Tc\ and \mcore\ do not differ, at the stage of evolution defined by a value of \xc, means that we may expect that the shift from the rotating to the non rotating configuration mainly implies a readjustment of the radius and luminosity, which become smaller, and the star moves to the location of  the \win=0 model, with its \xc\ value. Computation of specific models including braking is necessary, but the important point which will remain true is that a braked mass will find itself in {\it a less advanced stage of core--H consumption} with respect to the evolution at constant zero rotation rate. Notice that other physical phenomena which cause ingestion of hydrogen may remain active, e.g. those linked to the `overshooting' due to the finite velocity of convective elements at the convective borders.

\subsection{Data Availability Statement} The observational data published in this study are available from Antonino Milone upon reasonable request. The tracks, isochrone data and simulations which support the theoretical plots within this paper were retrieved by the authors from the SYCLIST (for SYnthetic CLusters, Isochrones, and Stellar Tracks), web facility available at \\
http://obswww.unige.ch/Recherche/evoldb/index/ and created by C. Georgy and S. Ekstr\"om. 

\begin{addendum}
 \item 
We thank C. Georgy and S. Ekstr\"om for creating and maintaining the interactive Web page for the Geneva stellar models at https://obswww.unige.ch/Recherche/evoldb/index/. AM acknowledges support by the Australian Research Council through Discovery Early Career Researcher Award DE150101816.

\item[Author Contributions]
F.D. and A.M jointly designed and coordinated this study. F.D. proposed and designed the rotational evolution model. F.D., E.V., A.M. and P.V. worked on the theoretical interpretation and implications of the observations. M.T. and M.D.C. carried out the simulations and the analysis. All authors read, commented on and approved submission of this article.

 \item[Competing Interests] The authors declare that they have no competing financial interests.
 \item[Correspondence] Correspondence and requests for materials
should be addressed to Francesca D'Antona~(email: francesca.dantona@oa-roma.inaf.it or franca.dantona@gmail.com).
\end{addendum}

~~~~~~

~~~~~~~~~

~~~~~~~~~~~~~

~~~~~

\textbf{SUPPLEMENTARY INFORMATIONS follow}

\newpage

\begin{figure}
\hskip -60pt
	\includegraphics[width=21cm]{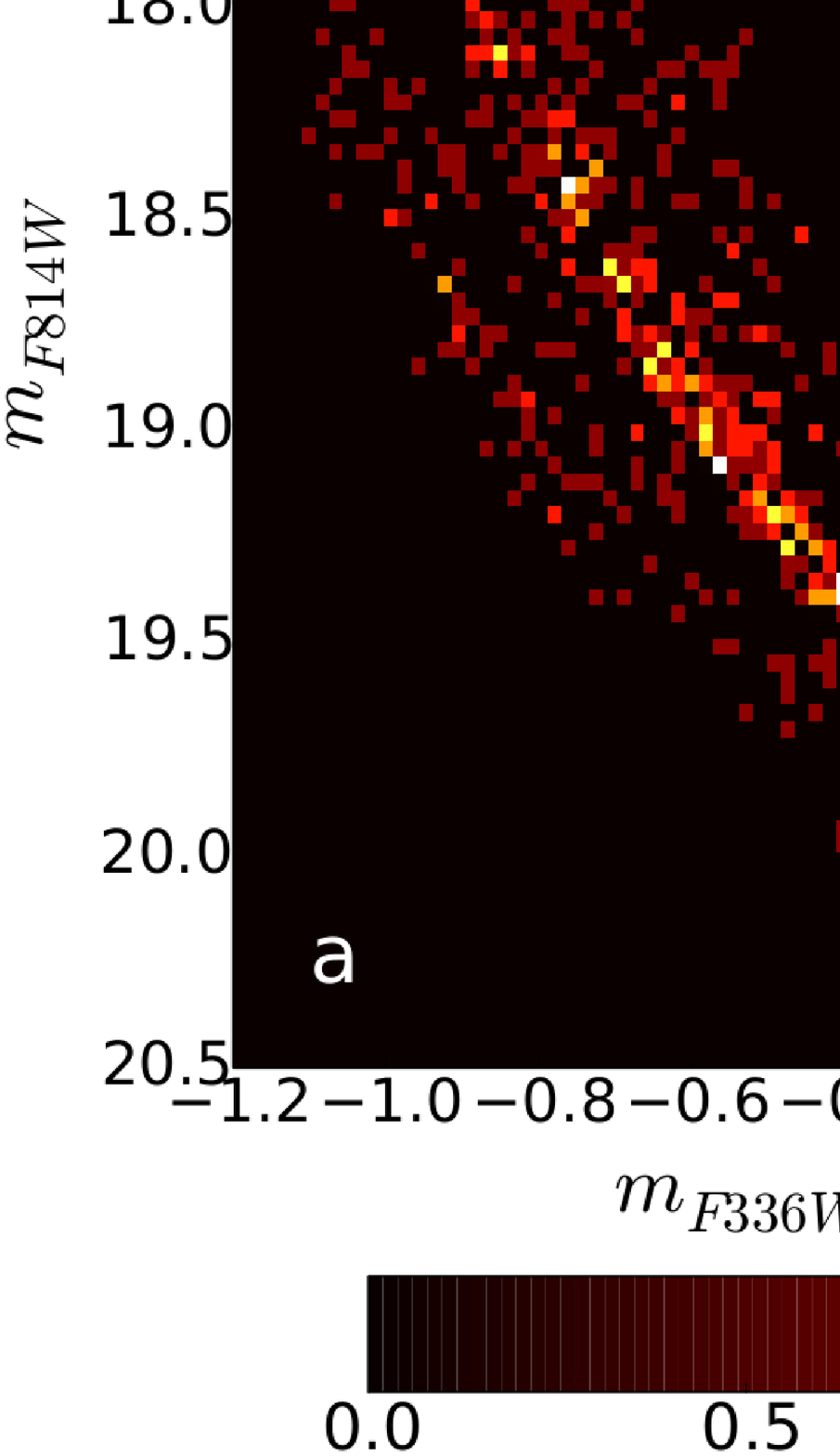}
\vskip +20pt  
    \caption{\textbf{Supplementary Figure~1~~---~~ Simulations for NGC\,1755 }
    The panels show the Hess diagrams for NGC\,1755 (color scale at bottom). \textbf{a}: data; \textbf{b}:  simulation with inclusion of a younger blue main sequence; \textbf{c}: simulation in which only the two coeval rotating+non rotating isochrones are employed. Population fractions of different stellar groups are listed in Supplementary table 1. }
    \label{EDfigure1}
\end{figure}
%%%%%%%%%%FIGURE 2 ED
\begin{figure}
\hskip -60pt
	\includegraphics[width=21cm]{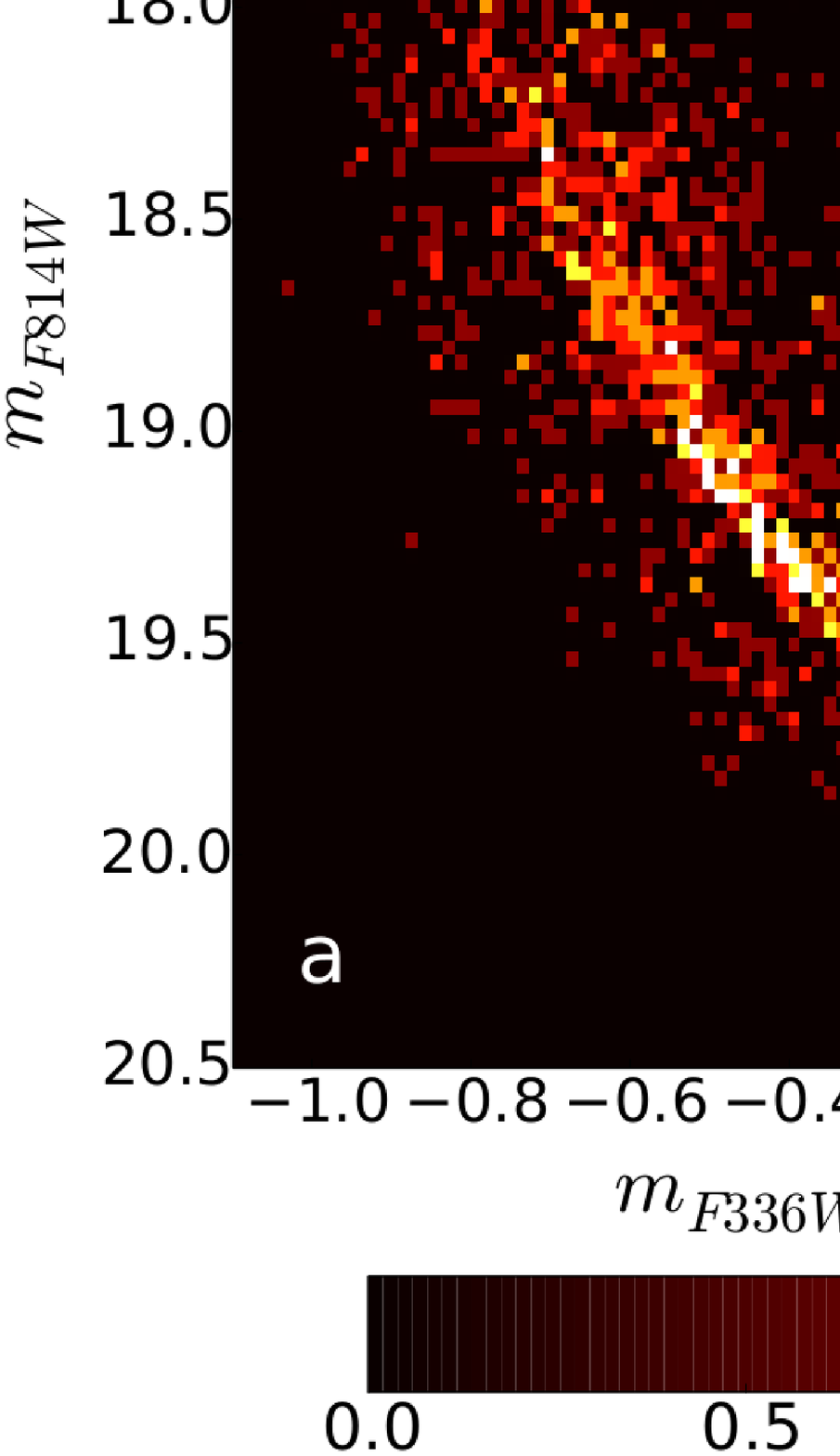}
\vskip +20pt  
    \caption{\textbf{Supplementary Figure~2~~---~~Simulations for NGC\,1850 }
The panels show the Hess diagrams for NGC\,1850 (color scale at bottom). \textbf{a}: data; \textbf{b}:  simulation with inclusion of a younger blue main sequence; \textbf{c}: simulation in which only the two coeval rotating+non rotating isochrones are employed.  Population fractions of different stellar groups are listed in Supplementary Table 1.  }
    \label{EDfigure2}
\end{figure}

\newpage
%%%%%%%%%%FIGURE 3 ED
\begin{figure}
%\vskip -60pt  
\hskip -60pt
	\includegraphics[width=21cm]{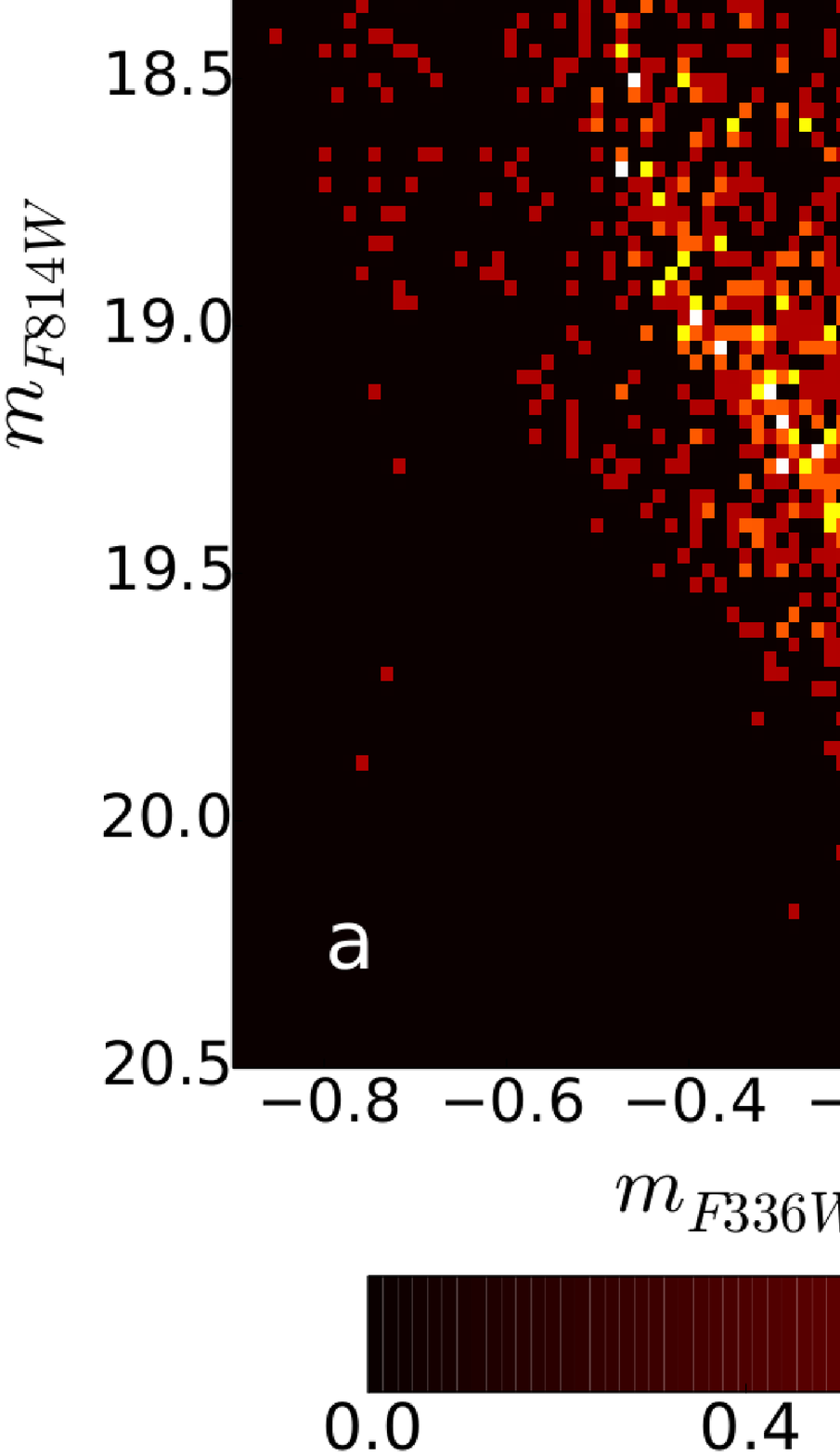}
\vskip +20pt  
    \caption{\textbf{Supplementary Figure~3~~---~~ Simulations for NGC\,1866 }
The panels show the Hess diagrams for NGC\,1866 (color scale at bottom). \textbf{a}: data; \textbf{b}:  simulation with inclusion of a younger blue main sequence; \textbf{c}: simulation in which only the two coeval rotating+non rotating isochrones are employed.  Population fractions of different stellar groups are listed in Supplementary table 1.  }
    \label{EDfigure3}
\end{figure}

%%%%%%%%%%FIGURE 1ADD. --NEW FIGURE 4
\begin{figure}
\vskip -20pt  
\center{
	\includegraphics[width=14.cm]{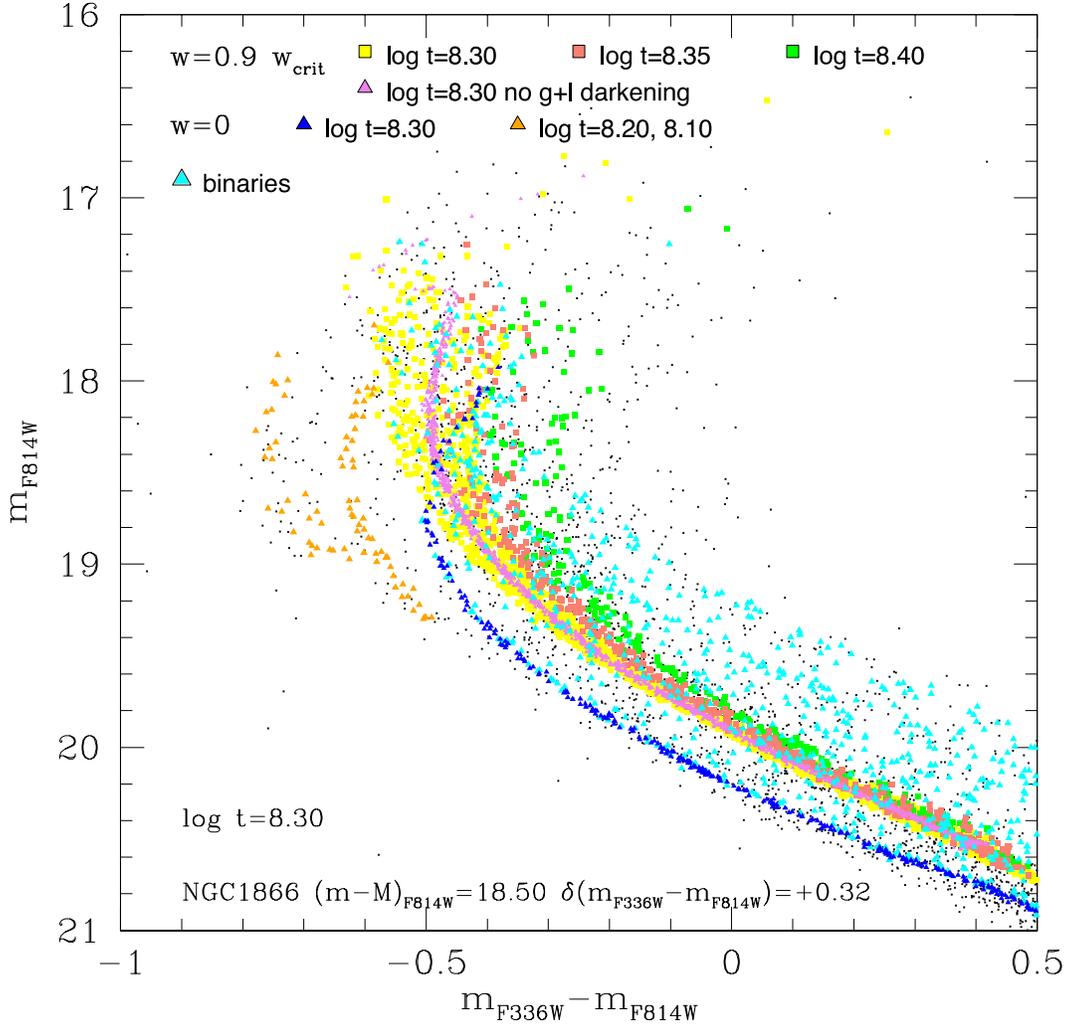}	
%\vskip -90pt  
    \caption{\textbf{Supplementary Figure~4~~---~~ Mapping the location of different stellar groups in NGC\,1866 }~~~The color magnitude data of NGC\,1866 (black small points, same as in Figure 1) are mapped by different stellar populations, labelled with their initial rotation $\omega$ and the logarithm of their age (in years). The numbers in each sample are not chosen to reproduce quantitatively the data, but to clarify the morphologies. The yellow and violet sequence are simulated from the same isochrone (rotating with $\omega$ equal to 0.9 times the break up angular velocity \wcrit);  but the violet points do not include projection effects (gravity and limb darkening, g+l in the label), showing that a part of the turnoff spread is due to rapid rotation and not age. The ``younger" non--rotating simulations and the ``older" rotating ones are added to represent the location of probably coeval stars, subject to the effect of fast braking as explained in the text. The bottom label shows the distance modulus and color shift (reddening) adopted for the simulated points. } }
    \label{EDfigure4}
\end{figure}

%%%%%%%%%%FIGURE 2ADD. --NEW FIGURE 5
\begin{figure}
\vskip -30pt  
\hskip -20pt
	\includegraphics[width=9.0cm]{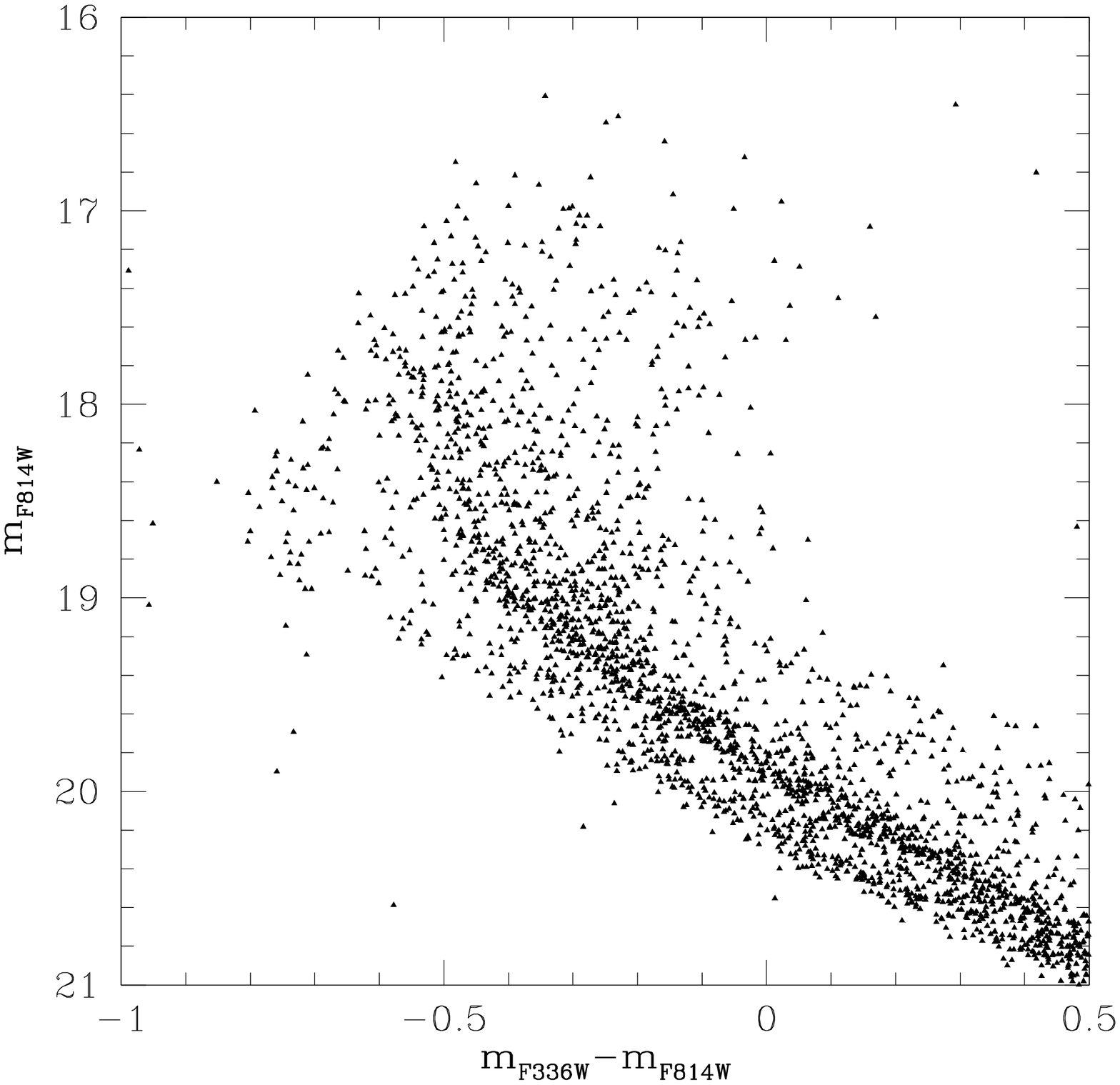}
	\includegraphics[width=9.0cm]{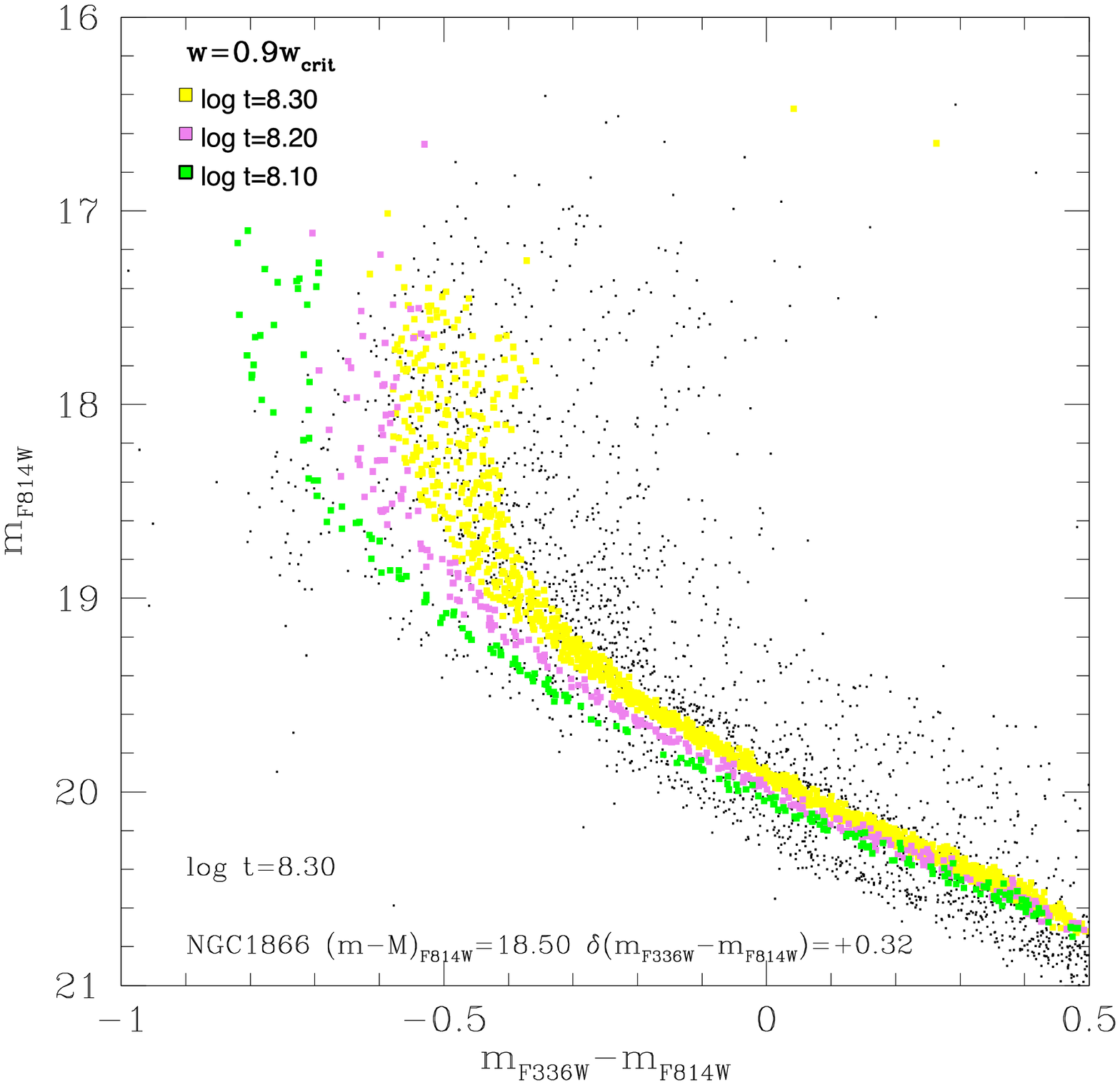}
%\vskip -50pt  
    \caption{\textbf{Supplementary Figure~5~~---~~  Why there are no rapidly rotating stars younger than the bulk of stars in NGC\,1866 }
    Color magnitude diagram of NGC\,1866 on the left, same data as in Figure 1. On the right, the simulation adopted to fit the red side of the diagram is shown in yellow, built on an isochrone of age 2$\times$10$^8$\,yr with initial angular velocity $\omega$ equal to 0.9 times the equatorial break up velocity \wcrit, and including projection effects by limb and gravity darkening. If ages younger by 0.1 and 0.2\,dex (the ages needed to fit the upper blue main sequence) are present, at the same rotation rate, the simulated samples populate the locus between the red and the blue side, where the observations do not show patterns of stars.  }
    \label{EDfigure5}
\end{figure}

%%%%%%%%%%FIGURE 4 ED. (6)
\begin{figure}
%\vskip -50pt  
%	\includegraphics[width=8.5cm]{./figurefinal/EDfigure1ab.pdf}
%	\includegraphics[width=8.5cm]{./figurefinal/EDfigure1cd.pdf}
	\includegraphics[width=8.5cm]{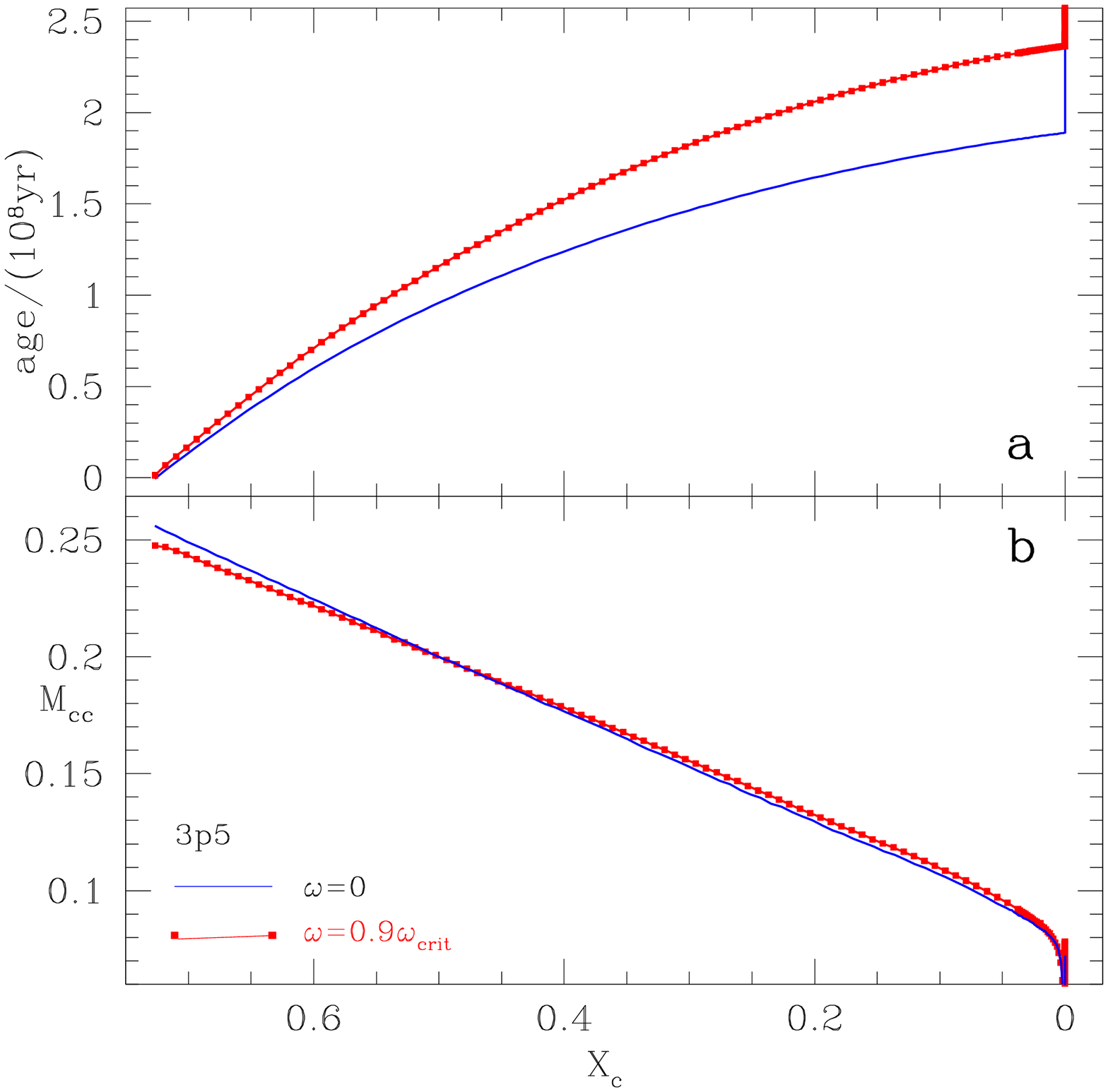}
	\includegraphics[width=8.5cm]{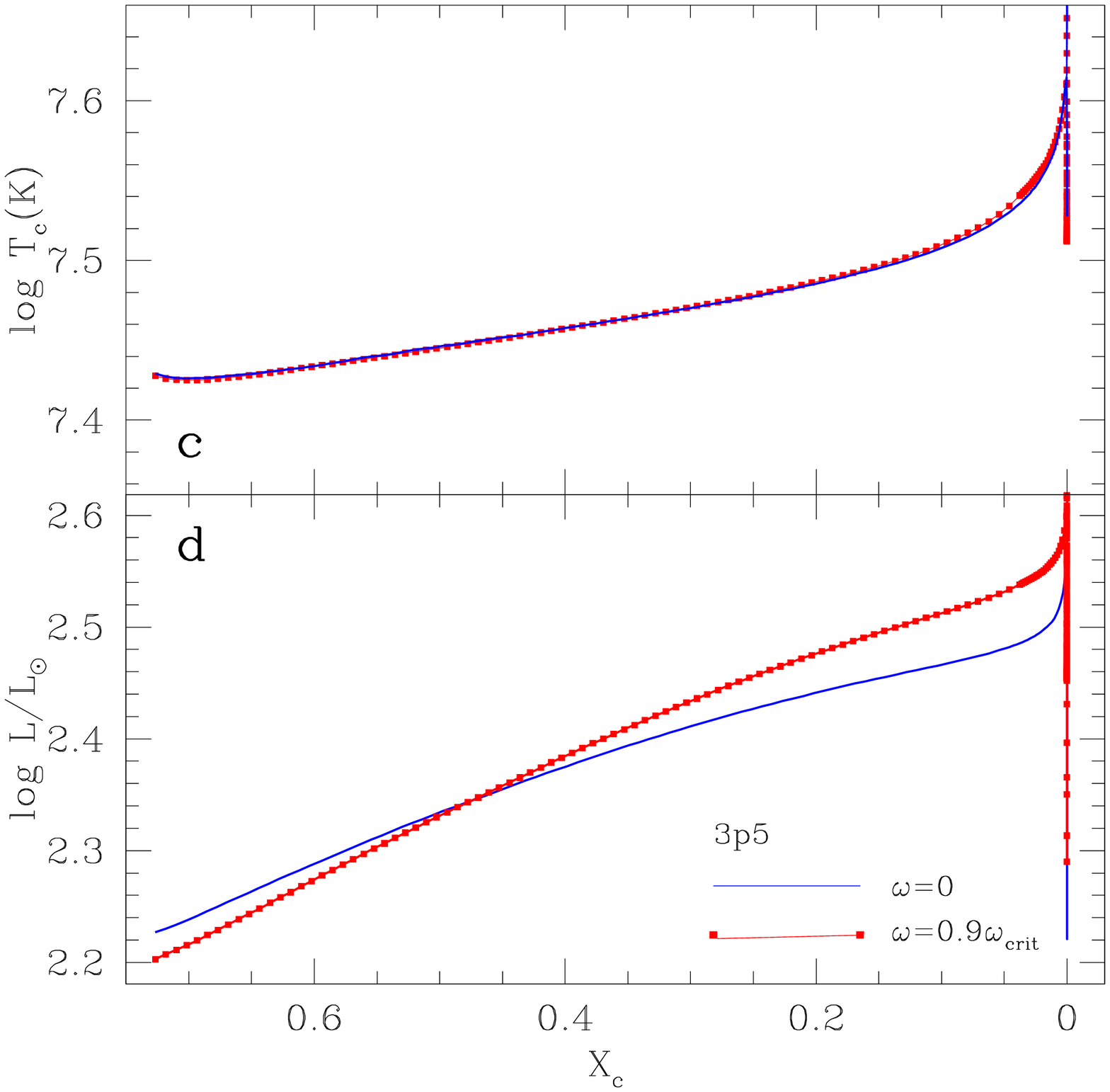}
 %\vskip -100pt  
	\includegraphics[width=8.5cm]{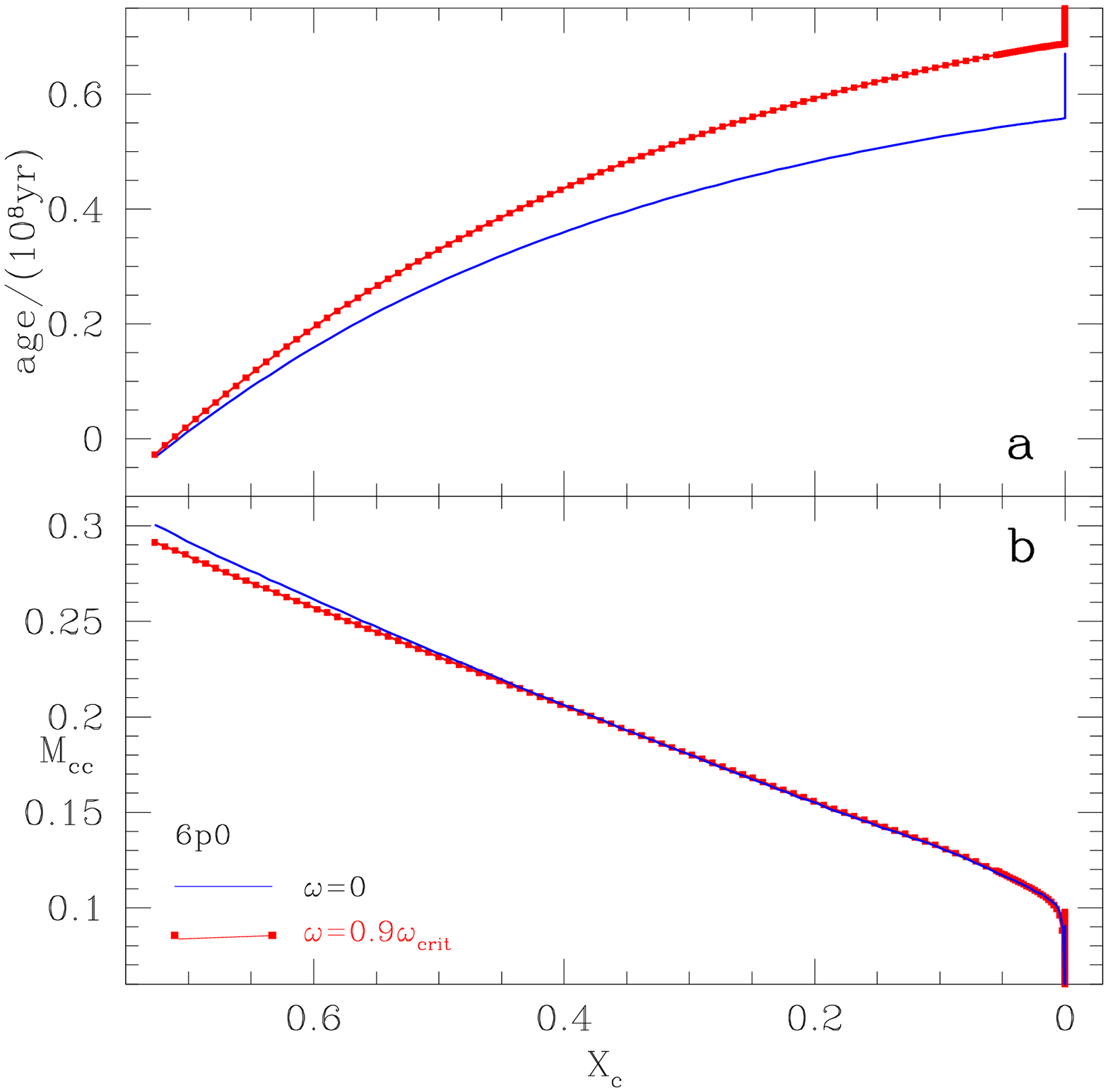}
	\includegraphics[width=8.5cm]{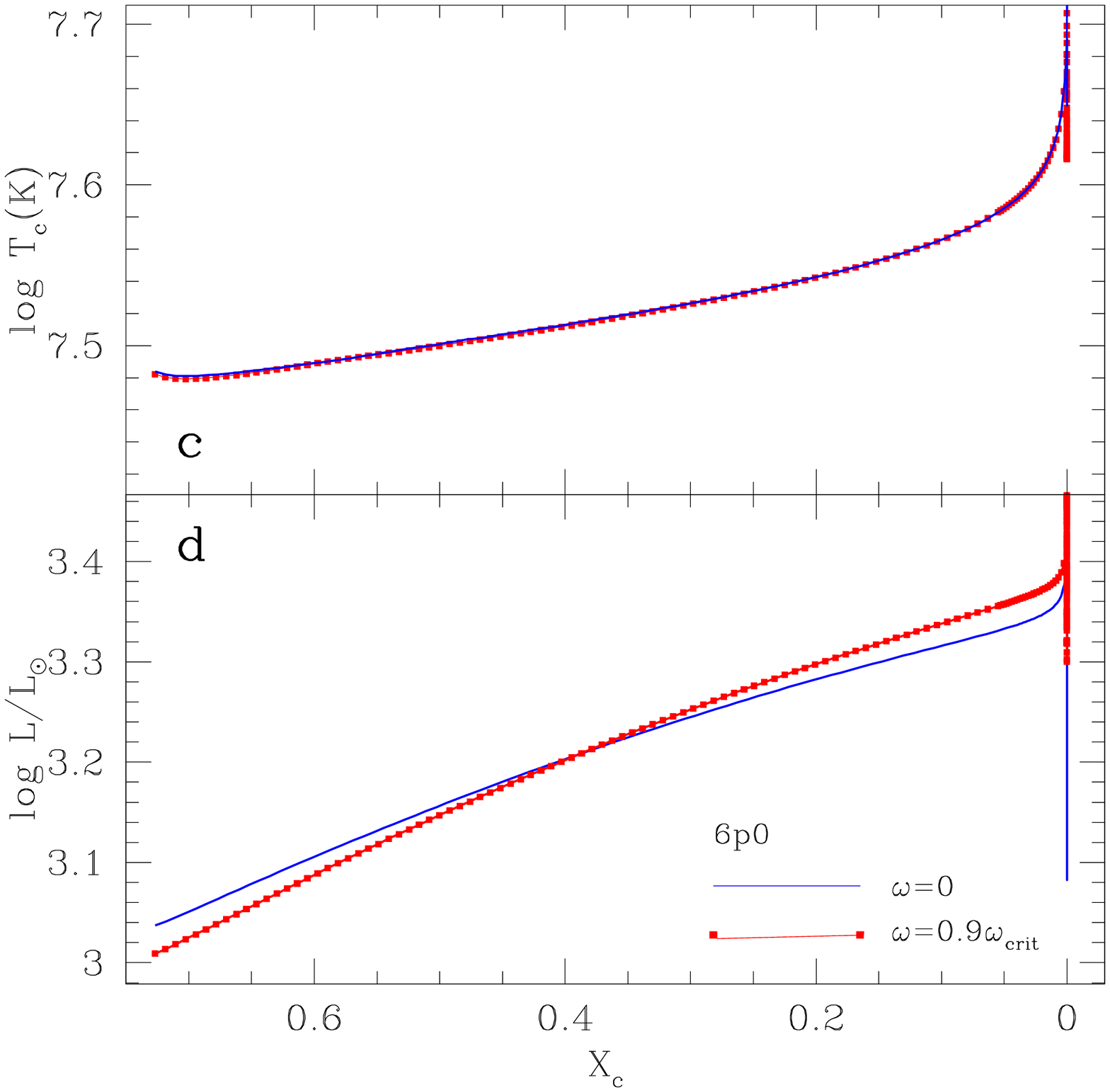}
 %\vskip -50pt  
 \caption{\textbf{Supplementary Figure 6 ~~---~~  Rotating versus non rotating core--H burning evolution for 3.5 and 6\Msun}
From the Geneva database, we show several physical quantities as a function of the core--H mass fraction \xc, for tracks rotating at angular velocity 0.9 the break up velocity \wcrit,  or non rotating.  a) time; b) convective core mass M$_{\rm cc}$; c) central temperature \Tc; d) luminosity. The four top figures refer to a mass M=3.5\msun\ (3p5), the bottom ones to M=6.0\msun\ (6p0). }
   \label{EDfigure6}
\end{figure}

%%%%%%%%%% ED Figure 5 (7)
\begin{figure}
\vskip -50pt  
	\includegraphics[width=18cm]{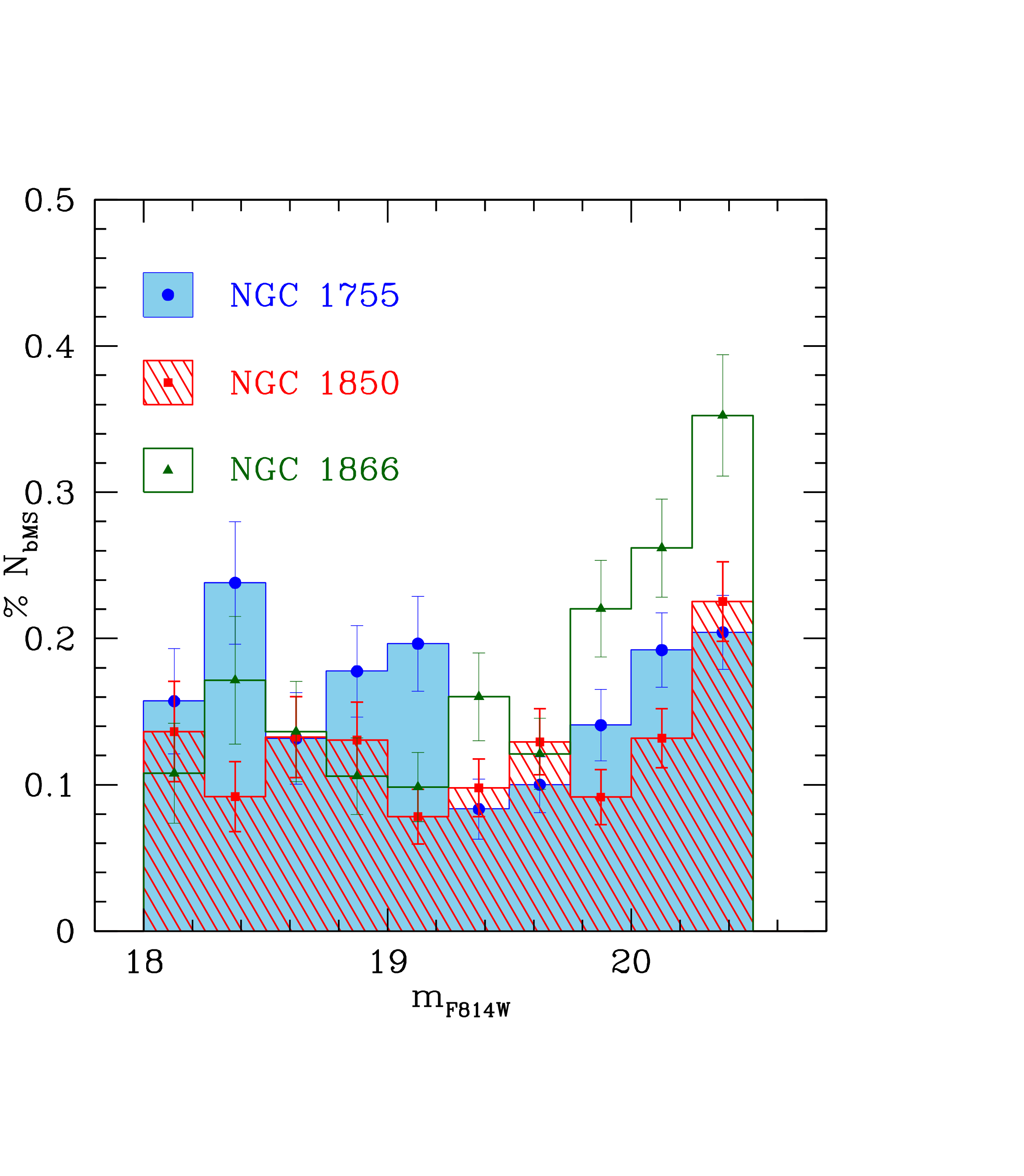}
%\vskip -60pt  
\caption{\textbf{Supplementary Figure~7 ~~---~~Ratio of blue main sequence stars to total, with error values, in NGC\,1755, NGC\,1850 and NGC\,1866.} The error in each bin has been calculated using standard error propagation starting from Poisson statistics. }
\label{EDfigure7}       
\end{figure}

\newpage

\begin{table*}
\caption{\textbf{Supplementary Table 1~~~Population fractions in the simulations of Supplementary Figure 1, 2 and 3}.\\
 The percentage of stars for each age group, rotating at angular velocity \win\ equal to 0.9 the equatorial break up velocity \wcrit, or not rotating, and the corresponding binary fractions are listed}.
\centering 
\begin{tabular}{c | c c | c c | c c}
\hline
 Cluster & \multicolumn{2}{c}{\win=0.9\wcrit  } & \multicolumn{2}{c}{\win=0.0} &  binaries & binaries   \\
             &  \%    & log(age/yr)            &      \%                 & log(age/yr)           &   \% \win=0.9\wcrit  &  \%  \win=0 \\
\hline
\hline
NGC1755          & 0.56 & 7.90  &  0.05  &  7.90  &   0.216   &  0.020 \\
                           &&                   & 0.11   & 7.80   &              &  0.044 \\
\hline
\hline
NGC1850     & 0.58 & 8.00  &  0.03  &  8.00    &  0.20 &  0.02 \\ 
                  &&           &    0.03    &    7.90  &   & 0.02   \\
                  &&           &    0.03    &    7.80  &    &  0.02  \\
                    & 0.04    & 8.10  &    &    &  0.03 &   \\ 
\hline
\hline 
NGC1866   & 0.55 & 8.30 &  0.05  & 8.30 & 0.20 & 0.016   \\
  &  &                  & 0.04  & 8.20 && 0.012\\
   &  &                 & 0.04   & 8.15 && 0.012\\
   &  &                 & 0.03   & 8.05 && 0.012\\
    &   0.03   & 8.35 & && 0.012& \\
    &  0.03   & 8.40 &  &&0.012 & \\
\hline
\end{tabular}
\label{TAB_POP}
\end{table*}

\newpage

\begin{table*}
\caption{\textbf{Supplementary Table 2: Stellar counts versus magnitude m$_{\bf F814W}$ for the clusters NGC\,1755, NGC\,1850 and NGC\,1866.} \\
Column 1: magnitude interval; for each cluster we list: $N_{\rm bMS}$, number of stars 
along the blue main sequence; $N_{\rm Tot}$: counts for all stars;  $N_{\rm bMS}/N_{\rm Tot}$: number ratio and error on the ratio.}
\begin{tabular}{c|rcc|rcc|rcc}
\hline
\hline
 & \multicolumn{3}{|c|}{NGC 1755} & \multicolumn{3}{|c|}{NGC 1850} &\multicolumn{3}{|c|}{NGC 1866}\\
\hline 
F814W$_{\rm In}$ & $N_{\rm bMS}$ & $N_{\rm Tot}$ & $N_{\rm bMS}/N_{\rm Tot}$ & $N_{\rm bMS}$ & $N_{\rm Tot}$ & $N_{\rm bMS}/N_{\rm Tot}$ & $N_{\rm bMS}$ & $N_{\rm Tot}$ & $N_{\rm bMS}/N_{\rm Tot}$ \\
\hline
18.00-18.25  & 11 & 70  & 0.16$\pm$ 0.04  &  18 & 132 & 0.14 $\pm$ 0.03  & 11 & 102 & 0.11 $\pm$ 0.03\\
18.25-18.50  & 20 & 84  & 0.24  $\pm$ 0.04   &  16 & 174 & 0.09 $\pm$ 0.02  & 18 & 105 & 0.17 $\pm$ 0.04\\
18.50-18.75  & 10 & 76  & 0.131 $\pm$ 0.03  &  26 & 196 & 0.13 $\pm$ 0.03  & 18 & 132 & 0.14 $\pm$ 0.03\\
18.75-19.00  & 19 & 107 & 0.178 $\pm$ 0.03  &  29 & 222 & 0.13 $\pm$ 0.03  & 18 & 170 & 0.11 $\pm$ 0.03\\
19.00-19.25  & 22 & 112 & 0.196 $\pm$ 0.02   &  19 & 243 & 0.08 $\pm$ 0.02 & 19 & 193 & 0.10 $\pm$ 0.02\\
19.25-19.50  & 9  & 108 & 0.083 $\pm$ 0.02 &  27 & 276 & 0.10 $\pm$ 0.02 & 33 & 206 & 0.16 $\pm$ 0.03\\
19.50-19.75  & 15 & 150 & 0.100 $\pm$ 0.02 &  37 & 286 & 0.13 $\pm$ 0.02  & 28 & 231 & 0.12 $\pm$ 0.02\\
19.75-20.00  & 19 & 135 & 0.141$\pm$ 0.02  &  26 & 284 & 0.09 $\pm$ 0.02 & 54 & 245 & 0.22 $\pm$ 0.03\\
20.00-20.25  & 34 & 177 & 0.192$\pm$ 0.03  &  48 & 364 & 0.13 $\pm$ 0.02 & 77 & 294 & 0.26 $\pm$ 0.03\\
20.25-20.50 & 39 & 191 & 0.204$\pm$ 0.03  &  84 & 373 & 0.23 $\pm$ 0.03 & 98 & 278 & 0.35 $\pm$ 0.04\\
\hline
\hline
\end{tabular}
\end{table*}

\end{document}